\documentclass[11pt,a4paper]{article}
\usepackage{amssymb,times,fullpage}
\usepackage{epsfig,theorem}

\newtheorem{theorem}{Theorem}
\newtheorem{lemma}{Lemma}[section]

\newtheorem{definition}{Definition}

\newcommand{\Xomit}[1]{}
\newcommand{\OPT}{\mbox{\textsc{opt}}}
\newcommand{\ALG}{\mbox{\textsc{alg}}}
\newcommand{\NF}{\mbox{\textsc{nf}}}
\newcommand{\A}{{\cal A}}
\newcommand{\B}{{\cal B}}
\newcommand{\I}{{\cal I}}

\newcommand{\R}{{\mathcal R}}

\newcommand*{\be}{\begin{equation}}
\newcommand*{\ee}{\end{equation}}
\newcommand{\ba}{\begin{eqnarray*}}
\newcommand{\ea}{\end{eqnarray*}}
\newcommand{\eps}{\varepsilon}

\newenvironment{proof}{\noindent {\bf Proof $\;$}}
                    {\hfill$\square$}

\begin{document}

\title{Improved results for a memory allocation problem}
\author{
Leah Epstein\thanks{Department of Mathematics, University of
Haifa, 31905 Haifa, Israel. {\tt lea@math.haifa.ac.il}. } 
\and
Rob van Stee\thanks{Department of Computer Science, University of Karlsruhe,
D-76128 Karlsruhe, Germany. \texttt{vanstee@ira.uka.de}. 
Research supported by Alexander von Humboldt-Stiftung. } } 
\maketitle


\begin{abstract}
We consider a memory allocation problem that can be modeled as a
version of bin packing where items may be split, but each bin may
contain at most two (parts of) items. A 3/2-approximation algorithm 
and an NP-hardness proof
for this problem was given by Chung et al.~\cite{ChGrVM06}.
We give a simpler 3/2-approximation algorithm for it which is in
fact an online algorithm. This algorithm 
also has good performance for the more general case where each bin 
may contain at most $k$ parts of items. We show that this general
case is also strongly NP-hard.
Additionally, we give an efficient 7/5-approximation algorithm.
\end{abstract}

\section{Introduction}

A problem that occurs in parallel processing is 
allocating the available memory to the processors. 
This needs to be done in such a way that each processor has
sufficient memory and not too much memory is being wasted.
If processors have memory requirements that
vary wildly over time, any memory allocation 
where a single memory can only be accessed by one processor
will be
inefficient. A solution to this problem is to allow memory
sharing between processors. However, if there is a single shared 
memory for all the processors, there will be much contention
which is also undesirable. It is currently infeasible to build
a large, fast shared memory and in practice, such memories
are time-multiplexed. For $n$ processors, this increases the
effective memory access time by a factor of $n$.

Chung et al.~\cite{ChGrVM06} studied this problem and 
described the drawbacks of the methods given above. Moreover, they
suggested 
a new architecture where each memory may be accessed by at most \emph{two} 
processors, avoiding the disadvantages of the two extreme earlier models.
They abstract the memory allocation problem
as a bin packing problem, where
the bins are the memories and the items to be packed represent
the memory requirements of the processors. This means that the
items may be of any size (in particular, they can be larger than
1, which is the size of a bin), and an item may be split, but
each bin may contain at most two parts of items. The authors of~\cite{ChGrVM06} 
give a $3/2$-approximation for this problem.

We continue the study of this problem and also consider a 
generalized problem where items can still
be split arbitrarily, but each bin can contain up to $k$
parts of items, for a given value of $k\geq 2$.

We study approximation algorithms in terms of the \textit{absolute
approximation ratio} or the \textit{absolute performance
guarantee}.
Let ${\B}(\I)$ (or $\B$, if the input $\I$ is clear from the
context), be the cost of algorithm $\B$ on the input $\I$. An
algorithm $\A$ is an $\R$-approximation (with respect to the
absolute approximation ratio) if for every input $\I$,
${\A}(\I)\leq \R\cdot {\OPT}(\sigma)$, where $\OPT$ is an optimal
algorithm for the problem. The absolute approximation ratio of an
algorithm is the infimum value of $\R$ such that the algorithm is
an $\R$-approximation.
The asymptotic approximation ratio for an online algorithm $\A$ is
defined to be
$$ \R^{\infty}_{\A} = \limsup_{n\to\infty}\sup_{\I}\left\{
{{\A}(\I) \over \OPT(\I)} \Bigg| \OPT(\I) = n\right\} \ .
$$

Often bin packing algorithms are studied using this measure. The
reason for that is that for most bin packing problems, a simple
reduction from the {\sc partition} problem (see problem SP12 in
\cite{GJ}) shows that no polynomial-time 
algorithm has an absolute performance
guarantee better than $\frac 32$ unless P=NP. However, since in our problem
items can be split, but cannot be packed more than a given number
of parts to a bin, this reduction is not valid. In
~\cite{ChGrVM06}, the authors show that the problem they study is NP-hard
in the strong sense for $k=2$. They use a reduction from the {\sc
3-Partition} problem (see problem [SP15] in \cite{GJ}). Their
result does not seem to imply any consequences with respect to
hardness of approximation. We show that the problem is in fact 
NP-hard in the strong sense for any fixed value of $k$.

A related, easier problem is known as bin packing with cardinality
constraints. In this problem, all items have size at most 1 as in
regular bin packing, and the items cannot be split, however there
is an upper bound of $k$ on the amount of items that can be packed
into a single bin. This problem was studied with respect to the
asymptotic approximation ratio. It was introduced and studied in
an offline environment as early as 1975 by Krause, Shen and
Schwetman \cite{KSS75,KSS77}. They showed that the performance
guarantee of the well known FIRST FIT algorithm is at most
$2.7-\frac{12}{5k}$. Additional results were offline
approximation algorithms of performance guarantee $2$. These
results were later improved in two ways. Kellerer and Pferschy
\cite{KP99} designed an improved offline approximation algorithm
with performance guarantee $1.5$ and finally a PTAS was designed
in \cite{CKP03} (for a more general problem). 

On the other hand,
Babel et al. \cite{BCKK04} designed a simple {\it online}
algorithm with asymptotic approximation ratio $2$ for any value of $k$. They
also designed improved algorithms for $k=2,3$ of asymptotic approximation
ratios $1+\frac{\sqrt{5}}{5}\approx 1.44721$ and $1.8$
respectively.  The same paper \cite{BCKK04} also proved an almost
matching lower bound of $\sqrt{2}\approx 1.41421$ for $k=2$ and
mentioned that the lower bounds of \cite{Yao80a,Vliet92} for the
classic problem hold for cardinality constrained bin packing as
well. The lower bound of 1.5 given by Yao \cite{Yao80a} holds for
small values of $k>2$ and the lower bound of 1.5401 given by Van
Vliet \cite{Vliet92} holds for sufficiently large $k$. No other
lower bounds are known.
Finally, Epstein~\cite{Epstein05} gave an optimal online bounded
space algorithm (i.e., an algorithm which can have a constant
number of active bins at every time) for this problem. Its
asymptotic worst-case ratio is an increasing function of $k$ and
tends to $1+h_\infty \approx2.69103$, where $h_\infty$ is the
best possible performance guarantee of an online bounded space
algorithm for regular bin packing (without cardinality
constraints). Additionally, she improved the online
upper bounds for $3\leq k \leq
6$. In particular, the upper bound for $k=3$ was improved to
$\frac 74$.

\paragraph{Our results}
In the current paper, we begin by showing that this problem is NP-hard
in the strong sense for any fixed value of $k$. This generalizes a result
from Chung et al.~\cite{ChGrVM06}. We also show that the simple NEXT
FIT algorithm has an absolute approximation ratio of $2-1/k$. This
matches and generalizes the performance of the more complicated
algorithm from~\cite{ChGrVM06}. 

Finally, we give an efficient $7/5$-approximation algorithm.


\section{NP-hardness of the problem (in the strong sense)}
\label{nphard} 

\begin{theorem}
Packing splittable items with a cardinality constraint of $k$
parts of items per bin is NP-hard in the strong sense for any
fixed $k\geq3$.
\end{theorem}
\begin{proof}
Given a fixed value of $k$, we show a reduction
from the 3-Partition problem defined as follows (see problem
[SP15] in \cite{GJ}). We are given a set of $3m$ positive numbers
$s_1,s_2,\ldots ,s_{3m}$ such that $\sum_{j=1}^{3m} s_j =mB$ and
each $s_i$ satisfies ${B\over 4}< s_i < {B\over 2}$.  The goal is
to find out whether there exists a partition of the numbers into
$m$ sets of size 3 such that the sum of elements of each set is exactly
$B$.  The 3-Partition problem is known to be NP-hard in the
strong sense. 

Given such an instance of the 3-Partition problem we define an
instance of the splittable item packing with cardinality
constraints as
follows. There are $m(k-3)$ items, all of size
$\frac{3k-1}{3k(k-3)}$ (for $k=3$, no items are defined at this
point). These items are called padding items. In addition, there
are $3m$ items, where item $j$ has size $\frac{s_j}{3kB}$ (for
$k=3$ we define the size to be $\frac{s_j}{B}$). These items are
called adapted items. The goal is to find a packing with exactly
$m$ bins. Since there are $mk$ items, clearly a solution which
splits items must use at least $m+1$ bins. Moreover, a solution
in $m$ bins contains exactly $k$ items per bin. Since the sum of
items is exactly $m$, all bins in such a solution are completely
occupied with respect to size.

If there exists a partition of the numbers into $m$ sets of sum
$B$ each, then there is a partition of the adapted items into $M$
sets of sum $\frac 1{3k}$ each (the sum is $1$ for $k=3$). Each
bin is packed with $k-3$ padding items and one such triple,
giving $m$ sets of $k$ items, each set of sum exactly $1$.

If there is a packing into exactly $m$ bins, as noted above, no
items are split and each bin must contain exactly $k$ items. If
$k=3$, this implies the existence of a partition. Consider the
case $k\geq4$. We first prove that each bin contains exactly $k-3$
padding items. 

If a bin contains at least $k-2$ padding items,
their total size is at least
$\frac{(3k-1)(k-2)}{3k(k-3)}=\frac{3k^2-7k+2}{3k^2-9k}=1+\frac{2k+2}{3k(k-3)}$.
For $k\geq 4$ this is strictly larger than $1$ and cannot fit
into a bin. If there are at most $k-\ell \leq k-4$ padding items,
then there are $\ell$ additional items of size at most
$\frac{1}{6k}$ ($\ell \geq 4$). The total size is therefore at
most
$\frac{(3k-1)(k-\ell)}{3k(k-3)}+\frac{\ell}{6k}=\frac{6k^2-2k-5\ell
k-\ell }{6k(k-3)}$. This value is maximized for the smallest
value of $\ell$ which is $\ell=4$. We get the size of at most
$\frac{6k^2-22k-4 }{6k(k-3)}=1-\frac{4(k+1)}{6k(k-3)}$. For $k
\geq 4$ this is strictly less than $1$, which as noted above does
not admit a packing into $m$ bins.

Since each bin contains exactly $k-3$ padding items, it contains
exactly three adapted items, whose total size is exactly
$\frac{1}{3k}$. The original sum of such three items is $B$, we
get that a solution in $m$ bins implies a partition.
\end{proof}


\section{The NEXT FIT Algorithm}

We can define NEXT FIT for the current problem as follows.
This is a straightforward generalization of the standard
NEXT FIT algorithm.
An item is placed (partially) in the current bin if the bin is
not full \emph{and} the bin contains less than $k$ item parts
so far. If the item does not fit entirely in the current bin,
the current bin is filled, closed, and as many new bins are
opened as necessary to contain the item.

Note that this is an online algorithm. The absolute approximation
ratio of NEXT FIT for the classical bin packing problem is $2$,
as Johnson~\cite{Johnso74} showed. Surprisingly, its
approximation ratio for our problem tends to this value for large
$k$. The two problems are different, and the two results seem to
be unrelated. 

Since items may be split, and we consider the
absolute approximation ratio,  this is the only reasonable online
algorithm that can be used for the problem.
We show that the approximation ratio of NEXT FIT is exactly 
$2-1/k$. Thus, this extremely simple algorithm performs as well as the
algorithm from~\cite{ChGrVM06} for $k=2$, and also provides the first
upper bound for larger values of $k$.

\begin{theorem}
The approximation ratio of NEXT FIT is $2-1/k$.
\end{theorem}
\begin{proof}
We first show a lower bound. The instance contains
an item of size $Mk-1$ followed by $M(k-1)k$ items of
size $\eps$, where $M$ is large and $\eps=1/(Mk(k-1))$.
Then the first item occupies $Mk-1$ bins, and the rest of the items
are $k$ per bin, in $M(k-1)$ bins. OPT has $Mk$ bins in total.
This proves a lower bound of $(M(2k-1)-1)/(Mk)$, which tends
to $2-1/k$ for $M\to\infty$.

Now we show a matching upper bound.

Let $u_1,u_2,\dots,u_m$  be sizes of the the blocks $1,\ldots,m$ of
NF. In each block, all bins are full except perhaps the last one, which
contains $k$ parts of items (except for block $m$, perhaps). We
assign weights to items. Let the size of item $i$ be $s_i$. Then
$w_i=\lceil s_i \rceil/k$. Note that in any packing, there are at
least $\lceil s_i \rceil$ parts of item $i$. Since there can be
at most $k$ parts in a bin, this means
\be
\label{eq:opt}
\OPT\geq\frac 1 k \sum_i \lceil s_i \rceil
=\sum_i \frac{\lceil s_i \rceil}{k}\ .
\ee
This explains our definition of the weights.
This generalizes the weight definition from Chung et al.~\cite{ChGrVM06}.

Consider the last bin from a block $i<m$. Since NF started a new
bin after this bin, it contains $k$ parts of items. Thus it
contains at least $k-1$ items
of weight $1/k$ (the last $k-1$ items are not split by the
algorithm). If $u_i=1$, there are $k$ such items. If $u_i>1$,
consider all items excluding the $k-1$ last items in the last
bin. We do not know how many items there are in the first $u_i-1$
bins (where the last item extends into bin $u_i$). 
However, for a fixed size $s$, the weight of a group of
items of total size $s$ is minimized if there is a single item in
the group (since we round up the size for each individual item to
get the weight). This implies the total weight in a block of
$u_i$ bins is at least $u_i/k+(k-1)/k=(u_i+k-1)/k$.

Now consider block $m$. If $u_i=1$, the weight is at least $1/k$
since there is at least one item. Else, as above the weight is at
least $u_i/k$, since the last bin of this block has at least one
item or a part of an item.

We have $\NF=\sum u_i$.
Therefore
\be
\label{eq:nf}
\OPT\geq  \sum_i w_i \geq \frac{\sum_{i=1}^m (u_i+k-1) -(k-1)}{k}
=\frac{\NF+(m-1)(k-1)}{k}\ .
\ee
Also by size, $\OPT > \NF - m$ and thus $\OPT \geq \NF-m+1$.
Multiply this inequality by $(k-1)/k$ and add it (\ref{eq:nf})
to get
$$\frac{2k-1}{k}\cdot\OPT \geq \NF\left(\frac 1 k+\frac{k-1}{k}\right)
+(m-1)\frac{k-1}{k}-(m-1)\frac{k-1}k
=\NF.$$
We conclude $\NF\leq (2-1/k) \OPT.$
\end{proof}

\section{The structure of the optimal packing for $k=2$}

Before we begin our analysis, we make some observations regarding
the packing of $\OPT$. A packing can be represented by a graph
where the items are nodes and edges correspond (one-to-one) to bins.
If there is a bin which contains (parts of) two items, there is an
edge between these items. A bin with only one item corresponds to
a loop on that item. The
paper~\cite{ChGrVM06} showed that for any given packing, it is possible
to modify the packing such that there are no cycles in the
associated graph. Thus the graph consists of a forest together
with some loops. 
We start by
analyzing the structure of the graph associated with the optimal
packing. Items of size at most $1/2$ are called \emph{small}.


\begin{lemma}
\label{ptas2-1}
There exists an optimal packing in which all small items are leaves.
\end{lemma}
\begin{proof}
Consider a small item that has edges to at least two other items.
Note that if two small items share an edge, the packing can be
changed so that these two items form a separate connected component
with a single edge. Thus we may assume that all neighbors are
(parts of) medium or large items.

Order the neighbors in some way and consider the first two neighbors.
Denote the small item by $s$ and the sizes of its neighboring parts
by $w_1$ and $w_2$. In bin $i$, $w_i$ is combined with a part $s_i$
of the small item $s$ $(i=1,2)$.

We have $s_1+s_2\leq1/2$. If $s_1\leq w_2$, we can cut off a part
of size $s_1$ from $w_2$ and put it in bin 1, while putting $s_1$
in bin 2. This removes neighbor $w_1$ from the small item $s$.

Otherwise, $w_2 < s_1\leq1/2$, which means that we
can put $s_1$ into bin 2 without taking anything out of bin 2:
we have $w_2 < 1/2$ and $s_1+s_2\leq1/2$. Again, $w_1$ is no
longer a neighbor of $s$ (or even connected to $s$).

Thus we can remove one neighbor from $s$.
We can continue in this way until $s$ has only one neighbor left.
\end{proof}
%




\begin{lemma}
\label{lem:typei}
An item of size in $((i-1)/2,i/2]$ has at most $i$ neighbors for all $i\geq2$.
\end{lemma}
\begin{proof}
Denote the items of size in $((i-1)/2,i/2]$ by type $i$ items.
We can consider the items one by one in each tree of the forest.

Consider a tree with at least one type $i$ item for some $i>1$ that
has at least $i+1$ neighbors. We want to create edges
between its neighbors and remove edges from the item
to the neighbors. However, these neighbors may be type $i$
items themselves, or some other type $j\geq1$.

We root the tree at an arbitrary item. Let the type of this
item be $i$. On this item
we apply the procedure detailed below. After doing this, the item
has an edge to at most $i$ other items. We define levels
in the tree in the natural way. Level 1 contains the root,
level 2 now contains at most $i$ items. We do not change
any edges going up from a particular level.

The items in level 2 undergoes the same procedure if necessary.
That is, if the number of its neighbors is larger than its type.
Afterwards, it only has $i$ neighbors, one of which is on level 1.
The other neighbors have moved to some lower level.

The procedure to remove a single neighbor of a type $i$ item
is as follows. For each item, we apply this procedure until
it has at most $i$ outgoing edges.
Consider a type $i$ item $x$ which is connected to at least $i+1$ other
items (generally: at least $i$ downlevel items).
Say part $m_j$ of item $x$ is with part $w_j$ of some other item in
bin $j$ for $j=1,\dots,i'$ where $i'>i$. 
If we are not dealing with the root
of the tree, let $w_{i'}$ be the uplevel node.

We sort the first $i'-1\geq i$ bins of this set in order of 
nondecreasing size of $m_j$. Since the total size of item $x$
is at most $i/2$, we then have $m_1+m_2\leq1$. These two parts can thus
be put together in one bin. This means cutting
one of the neighbors into two and moving it downlevel. We can do this
as long as the item has more than $i$ neighbors.
\end{proof}

\section{A $7/5$-approximation for $k=2$}
\label{sec:75}

Let $k=2$.
We call items of size 
in
$(1/2,1]$ \emph{medium} and remaining items \emph{large}. Our
algorithm works as follows. We present it here in a simplified
form which ignores the fact that it might run out of small items
in the middle of step 2(b) or while packing a large item in step
4. We will show later how to deal with these cases while
maintaining an approximation ratio of $7/5$. See Figure \ref{fig:75}.

\begin{figure} [hbt]
\begin{center}
\fbox{
\begin{minipage}{0.97\textwidth}
\begin{enumerate}
\item Sort the small items in order of increasing size, the medium
items in order of decreasing size, and the large items in order of
decreasing size.
\item Pack the medium items one by one, as follows, until you run
out of medium or small items.
\begin{enumerate}
\item If the current item fits with the smallest unpacked small item,
pack them into a bin.
\item Else, pack the current item together with the two \emph{largest}
small items in two bins.
\end{enumerate}
\item If no small items remain unpacked, pack remaining
medium and large items using Next Fit and halt. Start with the
medium items.
\item Pack all remaining small items in separate bins.
Pack the large items one by one into these bins using Next Fit
(starting with the largest large item and smallest small item).
\item If any bins remain that have only one small item, repack
these small items in pairs into bins and halt.
\item Pack remaining large items using Next Fit.
\end{enumerate}
\end{minipage}
}
\end{center}
\caption{The approximation algorithm for $k=2$}
\label{fig:75}
\end{figure}


We begin by giving an example which shows that this algorithm is
not optimal.
For some integer $N$, consider the input which consists of
$4N$ small items of size $2/N$, $2N$ medium items of size $1-1/N$,
$3N$ medium items of size $1-2/N$.

ALG packs the items of size $1-1/N$ in $4N$ bins, together with
$4N$ small items. It needs $3N(1-2/N)=3N-6$ bins for the remaining
medium items. Thus it needs $7N-6$ bins in total.

OPT places $3N$ small items in separate bins (one per bin),
and $N$ small items are split into two equal parts.
This gives $5N$ bins in which there
is exactly enough room to place all the medium items.

\begin{theorem}
\label{th:75}
This algorithm achieves an absolute approximation ratio of $7/5$.
\end{theorem}


The analysis has three cases, depending on whether the algorithm
halts in step 3, 5 or 6. The easiest case among these is without
a doubt step 5, at least as long as all bins packed in step 5
contain two small items.

\subsection{Algorithm halts in step 5}

%
Based on inequality (\ref{eq:opt}),
we define weights as follows.
\begin{definition}
\label{def:wt}
The weight of an item of size $w_i$ is $\lceil w_i \rceil/2$.
\end{definition}
In our proofs, we will also use weights of parts of items, based on
considering the total weight of an item and the number of its parts.
By Definition \ref{def:wt},
small and medium items have weight $1/2$. Therefore, we
have the following bounds on total weight of bins packed in the
different steps:
\begin{itemize}
\item[2.(a)] $1/2+1/2=1$
\item[2.(b)] We pack three items of weight $1/2$ in two bins, or
$3/4$ weight per bin on average.
\item[4.] Consider a large item which is packed in $g$ bins, that is,
together with in total $g$ small items. Its size is strictly
larger than $\frac{g-1}2$ and thus its weight is at least $g/4$.
Each small item has a weight of $1/2$, so we pack a weight of at
$3g/4$ in these $g$ bins.
\item[5.] $1/2+1/2=1$
\end{itemize}
This immediately proves an upper bound of $4/3$ on the absolute
approximation ratio.
There is, however, one special case: it can happen that one
small item remains unpaired in step 5. Since this case requires
deeper analysis, we postpone it till the end of the proof
(Section \ref{sec:step5}). 

\subsection{Critical items}

\begin{definition}
A \emph{critical} item is a medium item that the algorithm packs in Step 2(b).
\end{definition}

From now on, for the analysis we use a fixed optimal packing, denoted
by OPT. We consider the critical items 
in order of decreasing size. 
Denote the current item by $x$. We will
consider how OPT packs $x$ and define an \textbf{adjusted weight}
based on how much space $x$ occupies in the bins of OPT. Denote
the adjusted weight of item $i$ by $W_i$. The adjusted weights
will satisfy the following condition:
\begin{equation}
\label{eq:cond}
\sum_{i=1}^{n}\frac{\lceil w_i \rceil}{2}
\leq \sum_{i=1}^{n} W_i \leq \OPT.
\end{equation}
Specifically, we will have $W_i\geq \lceil w_i \rceil/2$ for $i=1,\dots,n$.
Thus the numbers $W_i$ will generate a better lower bound for $\OPT$,
that we can use to show a better upper bound for our algorithm.
This is the central idea of our analysis.
We initialize $W_i=\lceil w_i \rceil/2$ for $i=1,\dots,n$.
There are four cases.

\paragraph{Case 1}
\emph{OPT packs $x$ by itself.} 
In this case we give $x$ adjusted weight 1, and
so our algorithm packs an adjusted weight of 1 
in each of the (two) bins that contain $x$.
\paragraph{Case 2}
\emph{OPT packs $x$ with \emph{part} of a small item.}
Again $x$ and the bins with $x$ get an adjusted weight of 1.
This holds because when OPT splits a small item (or a medium item), 
it is as if it packs two small items, both of weight $1/2$.
Therefore such an item gets adjusted weight 1. We can transfer the
extra $1/2$ from the small item to $x$.

\paragraph{Case 3}
\emph{OPT combines $x$ with a \emph{complete} small item $y$.} 
Since our algorithm
starts by considering the smallest small items, $y$
must have been packed earlier by our algorithm, i.e.~with a larger
medium item $x'$ (which is not critical!).
If OPT packs $x'$ alone or with part of a small item,
it has an adjusted weight of 1 (Cases 1 and 2). Thus the bin with
$x'$ has an adjusted weight of $3/2$, and we transfer $1/2$ to $x$.
If OPT packs $x'$ with a full small item $y'$, then $y'$
is packed with a larger non-critical item $x''$ by our algorithm, etc.
Eventually we find a non-critical medium item $x^*$
which OPT packs alone or with
part of a small item, or for which Case 4 holds.
The difference between the weight and the adjusted weight of $x^*$
will be transferred to $x$. Note that the bin in which our algorithm
packs $x^*$ has a weight of 1 since $x^*$ is non-critical.
All intermediate items $x', x'',\dots$ have weight $1/2$ and are
non-critical as well, and we change nothing about those items.

\paragraph{Case 4}
\emph{OPT packs $x$ with a split medium or large item, or splits
$x$ itself.}

Since there might be several critical items for which Case 4 holds,
we need to consider how OPT packs all these items to determine
their adjusted weight.
We are going to allocate adjusted weights to items
according to the following rules:
\begin{enumerate}
\setlength{\itemsep}{0pt}
\item
Each part of a small item (in the OPT packing) gets adjusted weight $1/2$.
\item
A part of a large item which is in a bin by itself gets adjusted weight 1.
\item
A part of a large item which is combined with some other item
gets adjusted weight $1/2$.
\end{enumerate}

We do not change the weight of non-critical items.
The critical items receive an adjusted weight which corresponds to the number
of bins that they occupy in the packing of OPT. As noted above,
this packing
consists of trees and loops. Loops were treated
in Case 1. To determine the adjusted weights, 
we consider the \textbf{non-medium} items that are cut into parts by OPT.
Each part of such an item is considered to be a single item for this
calculation and has adjusted weights as explained above.
%
We then have that the optimal packing
consists only of trees with small and medium items, and loops. It can
be seen that each part of a non-medium item (for instance, part of
a large item) which is in a tree has weight $1/2$.

Consider a tree $T$ in the optimal packing. Denote the number of edges
(bins) in it by $t$. Since all items in $T$ are small or medium,
there are $t+1$ items (nodes) in $T$ by Lemmas \ref{ptas2-1} and
\ref{lem:typei}. Any items that are small
(or part of a small item) or medium but non-critical have
adjusted weight equal to the weight of a regular small or medium
item which is $1/2$. 
Denoting the number of critical items in $T$ by $c$, we find that the
$t+1-c$ non-critical items have weight $\frac{t+1-c}2$. 
All items together occupy $t$ bins in the optimal packing. This means
we can give the critical items each an adjusted weight of
$(t-\frac{t+1-c}2)/c = \frac12+\frac{t-1}{2c}$ 
while still satisfying (\ref{eq:cond}). This expression is minimized
by taking $c$ maximal, $c=t+1$, and is then $t/(t+1)$.
We can therefore assign an adjusted
weight of $t/(t+1)$ to each critical item in $T$.

Since the algorithm combines a critical item with two small items of
weight (at least) $1/2$, it packs a weight of $1+t/(t+1)=\frac{2t+1}{t+1}$
in two bins, or $\frac{2t+1}{2t+2}$ per bin.
This ratio is minimized for $t=2$ and is $5/6$.

However, let us consider the case $t=2$ in more detail. If the OPT tree with
item $x$ (which is now a chain of length 2) consists of three
critical items, then the sum of \emph{sizes} of these items is at
most $2$. Our algorithm packs each of these items with two small
items which do not fit with one such item. Let the sizes of the
three medium items be $m_1, m_2, m_3$. Let the two small items
packed with $m_i$ be $s_{i,j}$ for $j=1,2$. We have that
$m_1+m_2+m_3\leq 2$ but $m_i+s_{i,j}>1$ for $i=1,2,3$ and $j=1,2$.
Summing up the last six inequalities and subtracting the one
before, we get that the total size of all nine items is at least $4$.
Thus the area guarantee in these six bins is at least $2/3$.

If one of the items in the chain is (a part of) a small or large item,
or a medium non-critical item, it has adjusted weight $1/2$.
This leaves an adjusted weight of $3/4$ for the other two items. In
this case we pack at least $3/4+1=7/4$ in two bins, or $7/8$ per bin.
For $t\geq3$, we also find a minimum ratio of $7/8$.

Thus we can divide the bins with critical items
into two subtypes: $A$ with an adjusted weight
of $5/6$ and area $2/3$, and $B$ with an adjusted weight of
(at least) $7/8$ and area $1/2$.

\subsection{Algorithm halts in step 3}

\noindent We divide the bins that our algorithm generates into
types.
%
%
%
We have
\begin{enumerate}
\item groups of two small items and one medium item in two bins
\item pairs of one small item and one medium item in one bin
\item groups of four or more medium items in three or more bins
\item groups of three medium items in two bins
\item one group of bins with 0 or more medium items and all the large items
\end{enumerate}
Note that bins of type 4 contain a total weight of
at least $3/4$ ($3/2$ per two bins), as well as a total size of at
least $3/4$ (3 items of size more than $1/2$ in two bins). Thus,
whether we look at sizes or at weights, it is clear that these bins
can be ignored if we try to show a ratio larger than $4/3$.


Furthermore, in the bins of type 5 we ignore that some of the items
may be medium. The bounds that we derive for the total size and weight
packed into these bins still hold if some of the items are only medium-sized.

The bins of type 1 contain the critical items. We say the bins with subtype $A$ 
are of type $1a$, and the bins with subtype $B$ are of type $1b$.
Define $x_{1a},x_{1b,}x_2,x_3,x_4$ as the number of bins with
types $1a, 1b, 2, 3,$ and 5, respectively.

\Xomit{
Let $t\geq2$ be the size of the average group in type 4, rounded down.
Consider a group of size $t$, i.e., with $t+1$ items.
In such a group, the first $t$ items did not fit in $t-1$ bins,
and their average size is more than $(t-1)/t$. The largest size
in such a group is then also more than $(t-1)/t$. This is a lower
bound for \emph{any} item packed in Step 2 of the algorithm.
In particular, it is a lower bound for the size of an item in
a bin of types 1 and 2. (If there are no bins of type 4, we simply
take $t=2$.)
}

Consider the bins of type 3. Let $k$ be the number of groups of
medium items. Let $t_i \geq 3$ be the number of bins in group
$1\leq i \leq k$. The items in group $i$ have total size more than
$t_i-1/2$, since the last bin contains a complete medium item.
The total weight of a group is $\frac{t_i+1}{2}$, since it
contains $t_i+1$ items, each of weight $\frac 12$. We get that the total size of items in
bins of type 3 is at least $\sum_{i=1}^k (t_i-\frac 12)=x_3-\frac {k}{2}$, and the total weight of these items is
$\sum_{i=1}^k \frac{t_i+1}{2}=\frac{x_3+k}{2}$.


We find two different lower bounds on OPT.

Adjusted weight:
\be
\label{eq:wt1}
OPT \geq \frac 5 6 x_{1a} + \frac{7}{8}x_{1b} + x_2 + \frac{x_3}{2} +\frac k2 +
\frac{x_5}2 .
\ee
Size:
\be
\label{eq:si1}
OPT \geq \frac 2 3 x_{1a} + \frac{x_{1b}}2 + \frac{x_2}2 + x_3 - \frac{k}2  + \max(x_5-1,0).
\ee
%
%
Multiplying the first inequality by $\frac 45$ and the second one by $\frac 35$ we get
\be
\label{eq:joint}
\frac{7}{5} OPT \geq \frac {16}{15} x_{1a} + {x_{1b}} + \frac{11}{10} {x_2} + x_3 + \frac{k}{10}  +
\frac{2}{5}x_5+\frac{3}{5}\max(x_5-1,0).
\ee

If $x_5=0$ we are done. Else, (\ref{eq:si1}) is strict and we get
\be
\label{eq:si2}
OPT > \frac 2 3 x_{1a} + \frac{x_{1b}}2 + \frac{x_2}2 + x_3 - \frac{k}2  + x_5 - 1.
\ee
This means $x_3$ and $x_5$ occur with the same fractions in
(\ref{eq:wt1}) and (\ref{eq:si2}). Thus we can set $x_3:=x_3+x_5$ and
$x_5:=0$.
Adding (\ref{eq:wt1}) and (\ref{eq:si2}) and dividing by 2 gives
$$\OPT>\frac 3 4 (x_{1a}+x_2+x_3) + \frac{11}{16}x_{1b}-\frac 1 2\ .$$
This implies we are done if
$x_{1a}+x_2+x_3\geq \frac 3 4 x_{1b}+14.$
Clearly, this holds if any of $x_{1a},x_2$ or $x_3$ are at least 14.
Finally, by (\ref{eq:wt1}) we are also done if
$$\frac 5 6 x_{1a} + \frac{7}{8}x_{1b} + x_2 + \frac{x_3}{2} +\frac k2
\geq \frac 5 7 (x_{1a}+x_{1b}+x_2+x_3).$$
This holds if
$$\frac5{42} x_{1a} + \frac9{56} x_{1b}+\frac 2 7 x_2+\frac k 2
\geq \frac3{14}x_3.$$
Since we may assume $x_3<14$, we are in particular done if $x_{1b}\geq18$
or $k\geq 6$.

This leaves a limited set of options for the values of $x_{1a}, x_{1b},
x_2$, $x_3$ and $k$ that need to be checked.
It is possible to verify that for almost all combinations,
we find $\OPT\geq\frac 5 7 \ALG$.
One exception is $x_3=3$, $k=1$. However, going back
to the original variables, this means $x_3+x_5=3$ and $k=1$. But $x_3$
is either 0 or at least 3. If $k=1$, we must have $x_3=3$ and
$x_5=0$, so we
treated this case already.
%
%
%
%
%
Two other cases require special attention and are described below.

\Xomit{
Now, if $x_{1a}=x_2=0$, we can use that $x_{1b}$
is even (because it counts a number of bins that are packed in pairs) and
then from (\ref{eq:si2}) we get
$\OPT \geq x_{1b}/2 + x_3.$
Multiplying as above and adding gives
$$\frac{7}{5} OPT \geq x_{1b} + x_3$$
which is what we needed to show.

If $x_2\geq2$, we can subtract 2 from $x_2$ and add 2 to $x_{1b}$.
This does not affect $\ALG$ but decreases the bounds on $\OPT$.
Thus we can assume $x_2$ is 0 or 1.

Otherwise we have
$$\frac{7}{5} OPT > \frac {16}{15} x_{1a} + {x_{1b}} + \frac{11}{10} {x_2} + x_3 + \frac{k}{10}  +
x_5-\frac{3}{5} \ .$$
}


%
%
%
%

\paragraph{Special cases}

Step 2(b) requires two small items. If only one is left at this
point, and there is also no remaining medium item with which it could be
packed, we redefine it to be a medium item and pack it in step 3.
This leads to it being packed in a bin of type 3 (or 4).
Note that in this case, this small item and any
medium item we tried to pack with it in Step 2 have total size more than 1. Thus
if the small item ends up in a group of type 4 (a group of two bins), the total
size of the items in these bins (as well as the total weight) is still
at least 3/2, and we can ignore these bins in the analysis.
Therefore the analysis still holds.

There are two cases where $\OPT<\frac5 7 \ALG$ is possible.
If $x_2=1$ and $x_5=2$, a packing into two bins 
could exist in case there is only one large item. 
(If the bins counted in $x_5$ contain two medium
items, then we have that the three medium items require (at least)
two bins and the small item requires an extra bin.) If such a packing
exists, it works as follows: pack first the medium item,
then the large item (partially in the second bin), then the small
item. If this gives a packing into two bins, this is how our
algorithm packs the items. Otherwise we already have an optimal
packing.

If $x_{1b}=4$, $x_2=1$ and $x_5=5$, it is a simple matter to try
all possible packings for the items in 7 bins and check if one
is valid. (We can try all possible forests on at most 13 nodes
and at most 7 edges.)
If there is no packing in 7 bins, then our algorithm maintains the
ratio of $7/5$. If there is one, we use it.

\subsection{Algorithm halts in step 6}

In this case we have the following bin types.
\begin{enumerate}
\item groups of two small items and one medium item in two bins
\item pairs of one small item and one medium item in one bin
\item groups of large items with small items
\item one group of large items
\end{enumerate}
By definition, the type 1 bins contain the critical items. We
again make a distinction between type $1a$ bins with subtype $A$
and type $1b$ bins with subtype $B$.
In type 2 bins, the weight is 1.

Consider a large item which occupies 2 bins of type 3. This item
has weight of 1 and is combined with two small items in two
bins, giving a weight of 1 per bin. The large item also has size
more than 1, so an area of at least $1/2$ is packed per bin.
Comparing this to type $1b$ bins, which have a weight guarantee of
only $7/8$ but also an area guarantee of $1/2$, we find that we
may assume there are no such type 3 bins (with a large item
occupying two bins).

Now consider a large item which occupies 4 bins of type 3. Now we
find an overall weight of at least 3, as well as an overall size
of at least 3 (since the large item did not fit with 3 small
items in 3 bins). Since we plan to show a ratio larger than
$4/3$, we can ignore such bins as well. This also holds for large
items that occupy $g\geq5$ bins: the weight of the large item is
at least $g/4$ if $g$ is even and at least $(g+1)/4$ if $g$ is
odd.

We may therefore assume that all bins of type 3 form groups of
three bins, containing a weight of at least $5/2$ and an area
of at least 2. 
This gives a weight of $5/6$ per bin and an area
of $2/3$ per bin, just like type $1a$. We denote the number
of type $1b$ items by $x_1$ and the number of type $1a$ and
3 items by $x_3$.




Adjusted weight:
\be
\label{eq:wt6}
\OPT \geq \frac 7 8 x_1 + x_2 + \frac 5 6 x_3 + x_4/2.
\ee
Size:
\be
\label{eq:si6}
\OPT \geq \frac 1 2 x_1 + \frac 2 3 x_2 + \frac 2 3 x_3 +
\max(x4 - 1,0).
\ee
Multiplying and adding as in the previous section gives
$$\frac 5 7 \OPT\geq x_1+\frac 6 5 x_2+\frac{16}{15}x_3+\frac 2 5 x_4
+\frac 3 5 \max(x_4-1,0).$$
If $x_4=0$, we are done. Otherwise, we are done if
$\frac 1 5 x_2 + \frac 1{15}x_3\geq\frac 3 5$, which holds if
$x_2\geq3$ or $x_3\geq9$. Adding (\ref{eq:wt6}) and
(\ref{eq:si6}) gives that we are done if
$\frac 3 4 (x_3+x_4)\geq \frac 5 7 (x_3+x_4)+\frac 1 2$.
This holds in particular if $x_4\geq14$. Finally, from
(\ref{eq:wt6}) we get that we are done if
$\frac9{56}x_1\geq\frac3{14}x_4$, implying that we are done if
$x_1\geq\frac4 3\cdot 14$, or $x_1\geq19$. Again this
gives us a limited amount of choices to examine. Almost
all give us an approximation ratio of $7/5$.
The one exception to this case is $x_4=2$ and $x_2=1$, which we
can treat as in the previous section (repack into 2 bins if
possible). Other problematic cases, like ($x_4=2$ and $x_1=1$)
and ($x_4=2$ and $x_3=1$),
cannot occur because $x_1$ is even and $x_3$ is 0 or at least 2.

\paragraph{A special case: not enough small items to cover some large item}

If we run out of small items while packing some large item, this
large item is considered to be packed in step 6. That is, we
ignore the small items packed with this large item in our
analysis, and in the
last bin containing the large item we immediately continue with
the remaining unpacked large items. It can be seen that this does
not affect the weight or the area guarantee that we use for the
group of large items (indeed,
the weight guarantee improves somewhat, but we ignore this).

\subsection{One small item is unpaired in step 5}
\label{sec:step5}

By our analysis so far concerning the critical items, 
we know that bins packed in step 2(a) have
weight 1. Bins in step 2(b) are packed in pairs which have
adjusted weight at least $5/6$,
so $5/3$ per pair, although a pair only needs $10/7$ if we want to
show an approximation ratio of $7/5$. Bins in step
5 which contain a pair of small items have weight 1.

Thus if some items are packed in step 2(a) or 5 (as a pair),
we can transfer $1/4$ of adjusted weight to the bin with only one small item.
If a pair of bins is packed in step 2(b), we can transfer
$5/21$ of adjusted weight to the bin with the small item, which then has
more than $5/7$ of adjusted weight.

The only case left is where some bins are packed in step 4, and
one bin in step 5 (with one item). If there is a large item which
is packed into an odd number $g$ of bins, the weight of it is at least
$(g+1)/4$ and we are again done since we can transfer $1/4$. If $g$ is even and
at least 4, the weight is $g/4$.

If $g=2$, the weight of the large item is 1 and we find a weight of
1 per bin. So we may assume $g\geq4$ for all groups. This means that
all groups have an area guarantee of at least $3/4$.

Suppose all large items are packed into even numbers of bins.
Denoting the total number of bins
that we pack by $b$, we find that $b$ is odd (since there is exactly
one bin with only one small item) and that $b$ is equal to the
number of small items that we pack. The weight packed into these bins
is at least $(b-1)/4+b/2$. If $4|b-1$, this implies
$(b-1)/4+(b+1)/2$ bins are needed by OPT, which is more than $3b/4$.

If $4\nmid b-1$, there is at least one group of size 6 or more.
In this case we work with area guarantees:
the area guarantee in a group of size 6 is 5, and we find an area
guarantee of 5 for this group plus the lone bin with one small item,
or an area guarantee of $5/7$ per bin. (The remaining groups all
have area guarantee of at least $3/4$.)
This concludes the proof of Theorem \ref{th:75}.


\section{Conclusions}
In this paper, we gave the first upper bounds for general $k$ 
for this problem. Furthermore  we provided an efficient algorithm
for $k=2$.
An interesting question is whether it is possible to give an efficient
algorithm with a better approximation ratio for $k=2$ or for larger $k$.
In a forthcoming paper~\cite{EpsStexx} we will present approximation
schemes for these problems. However, these schemes are less efficient
than the algorithms given in this paper already for $\epsilon=2/5$.

\bibliographystyle{plain}


\end{document}